\documentclass{article}

\usepackage{graphicx}
\usepackage{fullpage,amsmath,amssymb,latexsym}
\usepackage{multirow}
\usepackage{natbib}
\begin{document}

\title{Core Formation in Giant Gaseous Protoplanets}
\author{Ravit Helled$^*$ and Gerald Schubert\\
\small{Department of Earth and Space Sciences and Institute of Geophysics and Planetary Physics,}\\
\small{University of California, Los Angeles, CA 90095 1567, USA}\\
\small{E-mail addresses: rhelled@ess.ucla.edu (R. Helled); schubert@ucla.edu (G. Schubert)}\\
\small{$^{*}${\it corresponding author}}\\
}

\date{}
\maketitle




\begin{abstract}
Sedimentation rates of silicate grains in gas giant protoplanets formed by disk instability are calculated for protoplanetary masses between 1 M$_{Saturn}$ to 10 M$_{Jupiter}$. Giant protoplanets with masses of 5 M$_{Jupiter}$ or larger are found to be too hot for grain sedimentation to form a silicate core. Smaller protoplanets are cold enough to allow grain settling and core formation. Grain sedimentation and core formation occur in the low mass protoplanets because of their slow contraction rate and low internal temperature. It is predicted that massive giant planets will not have cores, while smaller planets will have small rocky cores whose masses depend on the planetary mass, the amount of solids within the body, and the disk environment. The protoplanets are found to be too hot to allow the existence of icy grains, and therefore the cores are predicted not to contain any ices. It is suggested that the atmospheres of low mass giant planets are depleted in refractory elements compared with the atmospheres of more massive planets. These predictions provide a test of the disk instability model of gas giant planet formation. \\
The core masses of Jupiter and Saturn were found to be $\sim$ 0.25 M$_{\oplus}$ and $\sim$ 0.5 M$_{\oplus}$, respectively. The core masses of Jupiter and Saturn can be substantially larger if planetesimal accretion is included. The final core mass will depend on planetesimal size, the time at which planetesimals are formed, and the size distribution of the material added to the protoplanet. Jupiter's core mass can vary from 2 to 12 M$_{\oplus}$. Saturn's core mass is found to be $\sim$ 8 M$_{\oplus}$. 
\end{abstract}
\newpage

\section{Introduction}
The discovery over the last decade of extra-solar giant planets has increased interest in understanding the mechanism of giant planet formation (Lissauer \& Stevenson,~2007, Guillot, 2007). 
Currently, there are two models for the formation of gas giant planets: {\it'core accretion'}, the conventional model, and {\it'disk instability'}. The core accretion model suggests that the formation of a giant planet starts with planetesimal coagulation followed by accretion of a gaseous envelope. Once the core and envelope masses become equal, the rate of gas accretion increases significantly, leading to a hydrodynamic collapse of the atmosphere on top of the solid core (e.g., Pollack et al., 1996; Liassuer \& Stevenson, 2007). The disk instability model suggests that gas giant planets form as a result of gravitational fragmentation in the protoplanetary disk (e.g., Boss, 1997, 2007; Mayer, 2000, 2007). As self-gravitating clumps are created, dust grains can coagulate and sediment to form a core (Boss, 1997).  While in the core accretion model all giant planets are predicted to have rather similar core masses ($\sim$ 10 M$_{\oplus}$), giant planets formed by disk instability can, in principle, have no core. In the disk instability model, formation of a core is not a requirement, and the core mass would depend on the internal temperature of the protoplanet, the amount of available refractory material, the grain size distribution, and the strength of convection (see Helled et al., 2008 for further details).  In this paper we simulate the evolution of gas giant protoplanets created by disk instability and investigate how the core mass varies with the mass of the protoplanet.

\section{Model}
 \subsection*{Evolution}
It is still unknown whether objects formed by a gravitational instability can cool fast enough to remain bound objects that evolve to giant protoplanets (Rafikov, 2007). More detailed numerical simulations and further investigations are required before this question can be answered. For the purpose of this paper we assume that clumps formed by disk instability are gravitationally bound and contract further to form gaseous protoplanets.\\
To model the evolution of the protoplanets we assume gravitationally bound, spherical, homogenous and static bodies with a solar composition. The clumps are taken to be isolated, so no interaction with the surrounding disk is allowed. Under such conditions the bodies radiate away gravitational energy while their radii decrease. We concentrate on the initial stage in which the bodies are cold enough to allow the survival of silicate grains. This stage is known as the 'pre-collapse' stage  in which the protoplanets contract quasi-statically (DeCampli \& Cameron, 1979; Bodenheimer et al., 1980). The duration of this stage is inversely proportional to the initial mass of the body, so more massive protoplanets evolve faster. The pre-collapse stage ends when molecular hydrogen starts to dissociate at the center of the body and a hydrodynamic collapse ensues (DeCampli \& Cameron,1979; Bodenheimer et al., 1980).  

To model the evolution of protoplanets during the pre-collapse stage, we use a planetary evolution code that solves the standard equations of stellar evolution (hydrostatic balance, mass conservation, energy conservation, and heat transport). The code uses the interpolated equation of state of Saumon et al. (1995) for hydrogen and helium. The opacity is based on the work of Alexander \& Ferguson (1994) and the low-temperature grain opacity is based on the work of Pollack et al. (1985). The opacity includes both gas and grain, the latter based on the size distribution relevant for interstellar grains.  We find the initial quasi-static states based on a preliminary model of a simulated clump kindly provided
by P. Bodenheimer. Further details regarding the evolutionary model can be found in Helled et al.~(2006, 2008).\\
We follow the evolution of clumps with masses of 1 M$_{S}$, 1, 3, 5, 7 and 10 M$_{J}$ (M$_{S}$ and M$_{J}$ being the mass of Saturn and Jupiter, respectively). Internal temperatures decrease and the time-scale of the pre-collapse stage increases  with decreasing protoplanetary mass. The evolution of Jupiter and Saturn mass protoplanets is shown in Figure 1. The central pressure, central temperature, radius, and effective temperature as a function of time are presented. The evolution in the figures is shown up to the point when the central temperature reaches $\sim 1300$ K, the evaporation temperature of silicates, and core formation stops. The contraction rate of a Saturn mass protoplanet is slower, and its central pressure and temperature are lower compared with the Jupiter mass protoplanet. The pre-collapse time-scale of the low mass body is longer, and as a result grains inside it have a longer time to grow and sediment to the center.  

{\bf [Fig. 1]}\\

The initial states are chosen so that the protoplanets are extended and have low densities and temperatures and yet are gravitationally bound to insure further contraction (Boss,~1997; Mayer et al.,~2002). Table 1 presents the surface density, central pressure, central and effective temperatures, radius and luminosity of the initial states of the protoplanets.  More massive bodies can be more extended but they have higher internal temperatures. The surface density decreases for larger bodies but the central pressure, central temperature and central density increase with increasing mass. Massive bodies are hotter and larger and therefore are more luminous. Figure 2 presents the initial temperature-pressure profiles for protoplanets with 1, 5 and 10 Jupiter masses. As can be seen from the figure, the central temperature and pressure increase with increasing mass.\\

{\bf [Table. 1]}\\

{\bf [Fig. 2]}\\

The densities and temperatures of protoplanets formed by a gravitational instability are unknown, and different numerical models predict different characteristics. Simulations by Boss (1997, 2007) predict lower-density and colder protoplanets than the simulations of Mayer et al. (2002, 2004). The latter find denser and hotter configurations, with central temperatures reaching $\sim$1000 K shortly after formation. Indeed, for clumps with masses larger than $\sim$ 1 M$_{J}$ we find central temperatures of that order. The low surface densities of the protoplanets are a few times the mean density of the disk's midplane, as expected for bodies created by a gravitational instability (DeCampli \& Cameron, 1979). Though our choice of initial states is not well constrained, the evolution of the protoplanets depends mostly on their mass and is rather insensitive to the details of the initial states (DeCampli \& Cameron,~1979). Figure 3 presents the evolution of a Jupiter-mass protoplanet for three different initial states. The dotted curve represents a Jupiter mass object with an initial radius of $\sim$ 0.5 AU and a central temperature of 360 K. The solid and dashed-dotted curves show the evolution of Jupiter mass objects with initial radii of $\sim$ 0.3 and 0.2 AU and central temperatures of 560 and 800 K, respectively. Although the evolution of the objects is not identical, it is similar and physical parameters do not differ from each other by more than a factor of two. Differences are largest at the beginning of the evolution and decrease with time. Differences are found to be unimportant with regard to their effects on core formation.  
When the central temperature of the protoplanets reaches $\sim 1300$ K, dust grains evaporate and formation of the core stops. Protoplanets with masses of 5 M$_J$ or larger have central temperatures high enough to evaporate the grains even in their initial states and therefore they cannot form a core.

{\bf [Fig. 3]}\\
\subsection*{Sedimentation}
In order to examine whether core formation is possible, the sedimentation rate of the grains has to be computed.
The number of grains within the body can change due to the combined processes of coagulation and sedimentation, and is given by the Smoluchowski equation (Wetherill 1990),

$$\frac{\partial n(m,r,t)}{\partial t}={\frac 12}\int_0^m\kappa (m^{\prime
},m-m^{\prime })n(m^{\prime },r,t)n(m-m^{\prime },r,t)dm^{\prime
}$$
\begin{equation}
-n(m,r,t)\int_0^\infty \kappa (m,m^{\prime })n(m^{\prime
},r,t)dm^{\prime }-\nabla \cdot F.
\end{equation}
At each radius in the protoplanet $r$ at some time $t$ the number density of grains of mass between $m$ and $m+dm$ is given by $n(m,r, t)dm$.
$\kappa (m,m')$, the collision kernel, is the probability that a grain of mass $m$ will collide with and stick to a grain of mass $m'$.  Collisions occur due to the Brownian motion of the grains and the fact that larger grains with higher sedimentation speeds can overtake smaller grains. In convective regions, where different size grains respond differently to fluctuations in the gas, we use the approximation of Weidenschilling (1986), based on the work of Volk and collaborators (Volk et al. 1980; Markiewicz et al. 1991), which takes into account the relative velocities between different size grains in a turbulent gas.
We assume that whenever two grains meet and collide, they stick to form a bigger grain. The first integral gives the rate of formation of grains of mass $m$ while the second integral is equal to the
rate at which grains of mass $m$ are removed as a result of collisions with grains of any other mass. The third term on the right side $\nabla \cdot F$ is the 'transport term'. Transport of grains is possible due to sedimentation of grains (gravitational settling) or via turbulent transport in the convective regions. 
In the absence of convection the flux of grains of mass $m$ due to sedimentation is given by $F_{sed}(m,r,t)=n(m,r,t)v_{sed}(m,r,t)$ where $v_{sed}(m,r,t)$ is the sedimentation velocity of a grain of mass $m$ at planetary radius $r$ and time $t$.  The sedimentation velocity is found from the
force balance between the local gravitational force and the gas
drag on the grain (Podolak,~2003). In the case of convective transport, we use an
eddy diffusion approximation where the flux is given by $F_{conv}(m,r,t)=-K(r,t)\left[ \frac{\partial n(m,r,t)}{\partial
r}+\frac{n(m,r,t) }{H(r,t)}\right]$ where $H(r,t)$ is the pressure scale height and $K(r,t)$ is the eddy diffusion coefficient, which is given by
$K=vH$ where $v$ is the convective speed of the gas. The convective speeds are
estimated from the mixing length recipe (MLR).\\
Another effect that can change the grain distribution is grain sublimation, which occurs when the ambient temperature is high enough to evaporate the grains. Following Podolak (2003), we take the rate of vaporization to be
\begin{equation}
\frac{dm}{dt}=a^2P_{vap}(T)\sqrt{\frac{8\pi \mu}{N_AkT}},
\end{equation}
where $a$ is the grain's radius, $T$ is the grain's surface temperature, $\mu$ is the mean molecular weight of the grain material taken to be 50 (g/mole),
$P_{vap}=10^{\frac{-24605}{T}+13.176}$ (dynes cm$^{-2}$) is the vapor pressure of rocks (Podolak et al.,~1988), $k$ is Boltzmann's constant, and $N_A$ is Avogadro's number.
The surface temperature of the grain is taken to be the same as the temperature of the surrounding gas. As can be seen from equation 2, the evaporation rate is a function of the grain size. One can estimate the evaporation time as a function of temperature by computing $m/(\frac{dm}{dt})$ where $m$ is the grain's mass. Since the mass of a grain is proportional to $a^3$, the evaporation time is proportional to $a$, so larger grains require more time to evaporate. We take the evaporation temperature to be the temperature at which the evaporation time is shorter than $\sim$1 year. We find that for the grain sizes considered here (1, 0.1 and 0.01 cm), the evaporation temperature can differ by up to $\sim$ 100 K. We take the lower bound of this range and set the evaporation temperature to be 1300 K.\\
The grains are assumed to be solid silicate spheres with a density of 3.4 g cm$^{-3}$ and initial size $a_0$ and are initially homogeneously mixed within the protoplanet. 
The size distribution is divided among bins with radii a$_i$ that are logarithmically spaced in mass. The size of a grain in bin $i$, $a_i$, is taken to be $a_02^{i/3}$. The total number of bins is chosen to allow the grains to grow to sizes of at least 10 cm. Initially, only the smallest mass bin is populated, and as the grains collide and grow the larger size bins are filled.   
An object of solar composition is expected to have 2\% of heavy elements, but this mass also includes the volatile material (ices and organics). The mass of the refractory material is expected to be $\sim 1/3$ of the heavy element mass, and is taken to be  0.0067\% of the total mass. Re-condensation and breakup of grains are not included in the calculation. More details of the model can be found in Podolak (2003) and Helled et al.~(2008).\\

\section{Results}
Massive clumps can be extended and still be gravitationally bound but their internal temperatures are found to be too high to allow the grains to sediment and form a core. We find that silicate grains evaporate immediately in objects of 5 M$_{J}$ or larger, and instead of contributing to core formation, the grains evaporate in the planetary envelope enriching it with refractory material. Protoplanets with smaller masses are cold enough to let the grains survive. We follow the silicate grains as they grow and sediment towards the center of the protoplanet. In convective regions, the grains are carried by the convective eddies until they grow large enough to decouple from the gas and settle to the center. We find that once grains grow to sizes of 10 cm or larger (the exact value depends on the convection velocity which changes with depth, time and planetary mass), they are massive enough to decouple from the gas, sediment to the center and form a core. \\
The region of the core is represented by the innermost shell in the models, and its size varies with planetary mass. However, for all the cases considered here the size of this region is large ($\sim 10^{11}$ cm) and it contains gas in addition to the settled grains. Although this region might not represent a core in the traditional sense, the grains in this region remain there and they could eventually form a small silicate core in the center. The large number density of grains in the central region leads to frequent collisions that result in further growth. In addition, grains in the central region are typically larger than 10 cm, and therefore they are too large to be carried upward by convection into the planetary envelope. Once the central temperature is high enough to evaporate silicates ($\sim$ 1300 K) core growth stops. The refractory material inside the core region could segregate from hydrogen and helium and form an inner dense region. \\
Table 2 lists the derived core masses for different initial grain sizes (1, 0.1 and 0.001 cm) in protoplanets with masses between 1 M$_{S}$ to 10 M$_{J}$. Table 2 also gives the time available for core formation, i.e., the time it takes the body to reach internal temperatures of $\sim$1300 K. 

{\bf [Table. 2]}\\

We find that the core mass is \textit{not} simply proportional to the mass of the protoplanet. More massive protoplanets contain a larger amount of silicate grains but at the same time, the protoplanets are hotter and evolve faster. As a result, the available time to form a core decreases as protoplanet mass increases. Another effect that delays core formation in more massive bodies is the higher convective velocities. The convective flux is found to be more vigorous in massive bodies, so grains have to grow to larger sizes in order to decouple from gas and sediment to the center. Low-mass clumps have the lowest convective velocities and longest evolution - a configuration that supports core formation. However, due to their low mass the final core mass is relatively small. 
 
\noindent The resulting core masses for grains with initial radii between 0.01 cm and 1 cm are found to be very similar. This is due to the fact that small grains are coupled to the gas and have to grow to larger sizes (greater than $\sim$10 cm in radius) in order to to overcome the convective flux and settle to the center. Grains with smaller sizes than the ones considered here ($<$ 0.001 cm) are expected to result in similar core masses. The growth of small grains is very fast due to the large number density that results in frequent collisions and further growth (Helled et al.~2008). Grains with initial sizes larger then the sizes considered here would be large enough to quickly decouple from the gas and settle to the center. For these grain sizes, the core mass will be proportional to the planetary mass in protoplanets smaller than 5 M$_J$. However, such initial sizes seem unrealistically large for the early formation time of gaseous protoplanets by a gravitational instability (Boss, 1997). 
The last column in Table 2 gives the percentage of the core mass compared to the available grain mass. A Saturn mass body can sediment more than 80\% of the grain mass to its center, while more massive protoplanets sediment only $\sim 10\%$ of the available mass. 

\subsection*{Jupiter and Saturn}
A Saturn mass object is found to have the slowest contraction rate which leads to sedimentation of a larger amount of solids in comparison to a Jupiter mass object. In this study only  grains originally present in the body were considered, and the final core masses were found to be $\sim$ 0.25 and 0.5 M$_{\oplus}$ for Jupiter and Saturn, respectively.
In reality, protoplanets are embedded in the protoplanetary disk surrounding the young sun, and planetesimal accretion is likely to take place. Planetesimal accretion can add a significant amount of refractory material to the protoplanet that can settle to the center and increase the final core mass (Helled et al.,~2006; 2008).\\
Although conditions in the solar nebula are unknown, its characteristics can be estimated and therefore the planetesimal accretion rate of proto-Jupiter and proto-Saturn can be approximated. We assume that Jupiter and Saturn have not migrated significantly and define the total mass of solids in their feeding zones as (Pollack et al.~1996), 
\begin{equation}
M_{solids}=\pi(a_{out}^2- a_{in}^2)\sigma_{solids},
\end{equation}
where $a_{in}$ and $a_{out}$ are the inner and outer radii of the planetary feeding zone, respectively. The solid surface density of the disk is taken to be $\sigma=$10 g cm$^{-2}$ for Jupiter (5.2 AU) and $\sigma=3$ g cm$^{-2}$ for Saturn (9.5 AU), values that are a factor of $\sim$ 3 times higher than the "minimum-mass solar nebula" (MMSN) values (Pollack et al.,~1996). The planetary feeding zone represents the zone in which solid planetesimals are uniformly spread on either side of the orbit to a distance $a_f$, and its size changes with the eccentricity and inclination of the planetesimals' orbits and the planetary Hill sphere radius (Pollack et al.~1996). $a_{out}=a+a_f$ and $a_{in}=a-a_f$, where $a$ is the planet's semimajor axis. Following Pollack et al.~(1996), we take $a_f\sim$2$\times10^{13}$ cm and $a_f\sim$2.6$\times10^{13}$ cm for Jupiter and Saturn, respectively.  
The size of the feeding zone is taken to be constant with time, so planetesimals cannot get into or out of the feeding zone. Under these assumptions, the mass of solids in Jupiter's and Saturn's feeding zone are found to be $\sim 35$ M$_{\oplus}$, and $\sim 23$ M$_{\oplus}$, respectively. \\

For simplicity, we assume that a similar percentage of the grains sediments to the core as in the case when only the grains originally present in the body were considered (previous section). In principle, when planetesimals are accreted they lose material due to gas drag as they pass through the planetary envelope until they break or get fully evaporated. As a planetesimal penetrates deeper into the interior, the surrounding density increases, the drag force is stronger, and more material can evaporate. In that case the additional silicates will tend to be more concentrated in central regions. Our assumption, that the additional material is distributed in proportion to the gas density, results in a concentration of silicates toward the center (Helled et al., 2008). \\

The planetesimals are assumed to be composed of a mixture of ice,
silicates and CHON (an average density of $\sim2$ g cm$^{-3}$). Again, since the solid material contains ices and organics as well as silicates, the available rocky material is smaller ($\sim$1/3 of the total mass) and is taken to be $\sim$ 12 M$_{\oplus}$ and 8 M$_{\oplus}$ for Jupiter and Saturn, respectively.
The final core masses of Jupiter and Saturn can change with planetesimal size, the time in which planetesimals are formed, and the size distribution in which the additional material is added to the protoplanets. Three possible scenarios are considered:
\begin{enumerate}
\item \textit{Planetesimals are present during the pre-collapse stage and the additional material is added as small grains:} 
For planetesimal sizes of 10 km (or smaller), we find that the long contraction times of the protoplanets insure the accretion of all the solid material in their feeding zone before the internal temperatures become high enough to evaporate the rocky material. Taking the accreted material added to the protoplanet as small grains with similar sizes as the ones originally present in the body, Saturn's core is found to be as big as 8 M$_{\oplus}$, while Jupiter's core mass does not exceed 2 M$_{\oplus}$. 
\item \textit{Planetesimals are present during the pre-collapse stage and the additional material is added as planetesimal fragments:}
There is a good chance that the size distribution of the additional material is quite different from that of the initial grain distribution in the envelope. As a planetesimal passes through the planetary envelope pressure gradients across the planetesimal
can exceed the material strength of the body, resulting in planetesimal fragmentation (Podolak et al.,1988). Although some fraction of the planetesimal mass will be evaporated in the envelope as the planetesimal passes through the gas, most of the mass will be in fragments of a few meters in size. Fragments with these sizes have high sedimentation speeds and will settle to the center almost immediately. Most of the accreted mass will then be in the core region. The final core masses of Jupiter and Saturn can then be as large as $\sim$ 12 M$_{\oplus}$ and 8 M$_{\oplus}$, respectively.  
\item \textit{Planetesimals are not present during the pre-collapse stage:}
Gravitational instabilities occur relatively early in the history of the protoplanetary disk ($\sim10^3$ years) and it is not clear whether planetesimals have been formed (Boss,~1997; Dominik et al., 2007). If planetesimals are not yet formed, there will be no additional material contributing to core formation and the core masses of Jupiter and Saturn will be as small as 0.25 M$_{\oplus}$ and 0.5 M$_{\oplus}$, respectively (Table 2). The planetesimals could be accreted at a later stage of the planetary evolution when the protoplanets are significantly hotter. In this case, the accreted material will contribute to the planetary envelope enrichment in silicates, ices and organics. In this scenario Jupiter and Saturn core masses are too low compared to theoretical models that estimate Jupiter and Saturn core masses (Saumon \& Guillot, 2004). 
\end{enumerate}

\section{Discussion and Conclusions}
We followed the evolution of gaseous protoplanets formed by disk instability. We considered protoplanets with masses between 1 Saturn mass and 10 Jupiter masses, a mass range that covers most of the observed mass range of extra-solar giant planets. The protoplanets were taken to be extended and yet gravitationally bound to insure further contraction. We examined the possibility of core formation as a result of sedimentation of silicate grains. Grain settling can occur in protoplanets which are cold enough and have a contraction time-scale long enough for the grains to grow (decouple from the convective flux in convective regions) and sediment to the center. The grain sedimentation process is found to be favorable for low mass bodies, due to lower internal temperatures, lower convective velocities and longer contraction time in the pre-collapse stage. We find that protoplanets with masses of $\geq$ 5 M$_{J}$ are too hot to allow core formation and therefore will have no solid cores. \\
The core masses presented in Table 2 are given under the assumption that the protoplanets are isolated and the grain mass is taken to be a small fraction ($\sim 0.007\%$) of the planetary mass. As discussed earlier, planetesimal accretion can add a significant amount of refractory material to the protoplanets (Helled et al., 2006,2008). However, the accreted material contributes to core formation only if accretion occurs before the internal temperatures are high enough to evaporate the grains. Only protoplanets with masses lower than 5 M$_J$ will have larger cores than presented in Table 2. In the case of more massive protoplanets, the additional solids would only enrich the planetary envelope with silicates, ices and organics regardless of the accretion timing. If a significant amount of material is accreted during the pre-collapse stage, the core mass in low mass protoplanets is expected to be inversely proportional to the planetary mass. The final core mass will not only depend on the time at which the solids are captured but also on the solid surface density of the protoplanetary disk, the distance at which the planet is formed, etc. We estimated the final core masses of Jupiter and Saturn when planetesimal accretion was included. However, due to uncertainties in the size distribution of the accreted material as it enters the protoplanet, a range of core masses is found. Jupiter and Saturn core mass can change from 2 to 12 M$_{\oplus}$, and $\sim$ 8 M$_{\oplus}$, respectively (see previous section). \\
The results presented here are most robust for protoplanets with masses $\geq$ 5 M$_J$.  The temperature gradient inside these bodies is least effected by stellar heating, and in addition, the core size is unaffected by planetesimal accretion, since these planets have no core due to their high internal temperatures. If heavy elements are added due to planetesimal accretion they will be spread throughout the planet's interior. Our conclusion that massive giant planets have no solid cores differs from the core accretion model which predicts that massive giant planets have a solid core of the order of $\sim$ 10 M$_{\oplus}$ (Pollack et al.,~1996; Hubickyj et al.,~2005). Future determination of the core mass of massive extra-solar giant planets could help to distinguish between the two possible mechanisms for giant planet formation. \\
Gravitational settling of dust is also known to occur in brown dwarf atmospheres. Grain sedimentation explains the element depletion, spectra and weather-like features in cool brown dwarf atmospheres (Woitke \& Helling, 2002). In a similar process as considered here, drops of dust with sizes large enough to overcome the convective flux settle below the visible atmospheres, forming cloud layers (Ackerman \& Marley, 2001; Helling \& Woitke, 2006). Grain sedimentation (and cloud formation) can therefore, also occur in a later stage of the evolution of the bodies considered here. However, in this case the dust depletion in the atmosphere will not be a result of core formation, but grain settling below the observed atmosphere. This process imparts further complexity to observational tests of the theory. \\
Our results rely on several assumptions. The protoplanets were taken to be isolated so radiation from the young star was not included. Clumps forming in planetary disks are exposed to stellar radiation, especially in the inner part of the disk. Stellar radiation can heat the bodies, increasing the internal temperatures and inhibiting core formation. To keep the calculation as general as possible we chose to make no assumptions regarding the location of the clump. In addition, since in the disk instability model protoplanets are usually formed in the cold outer regions where the disk is cold enough to become gravitationally unstable (Boss,~1997; Mayer et al., 2002) stellar radiation is expected to be relatively weak. Nevertheless, the effect of stellar radiation should be included in future work, especially due to the possibility of inward migration that can place the bodies in regions where stellar radiation is stronger (Papaloizou et al., 2007).\\  
The core formation process itself was not included in the evolutionary simulation. Grain settling can generate additional luminosity in the protoplanet with an energy release of $\sim$ $\frac{GM_c^2}{R_ct}$,
where $G$ is the gravitational constant, $M_c$ is the core mass, $R_c$ is its radius and $t$ is the time for core formation.
For the core masses and radii considered in this work, the luminosity produced by core formation was found to be small in comparison to the luminosity due to planetary contraction (Table 1). The effect of core formation on the planetary evolution for the cases considered here is found to be small. However, formation of a massive compact core can increase the luminosity proportionally, and the additional energy can become comparable to or even larger than the luminosity generated by gravitational contraction. In that case it would be necessary to include the additional energy in the planetary evolution.\\ 
\indent Grains composed of ice or complex organics (CHON) were not considered in the present work. These materials are more volatile than silicates and evaporate at lower temperatures. Even the coldest model we have considered is found to be too hot to allow the survival of CHON and icy grains. As a result, cores of giant planets created by a gravitational instability are expected to be composed only of refractory material. We find that even a Saturn-mass protoplant is too hot to allow the survival of these volatiles. We conclude that the formation of icy planets via gravitational instability followed by photoevaporation of their gaseous envelopes as suggested by Boss et al. (2002) is rather unlikely. \\
\indent Our results lead to an interesting conclusion regarding the composition of the planetary envelope. We find that the percentage of the core mass increases with decreasing total mass. As a result, the atmosphere of low mass giant planets formed by a gravitational instability is expected to be depleted in refractory material compared to more massive planets ($\geq$ 5 M$_J$). Measurements of the composition of extrasolar planetary atmospheres could reveal important information regarding the formation mechanism of the giant planets. \\

\noindent{\bf Acknowledgments: }\\
\noindent The authors thank L. Mayer for inspiring discussions and valuable information.
\section{References} 
\small{
Ackerman, A. S. \& Marley, M. S., Precipitating Condensation Clouds in Substellar Atmospheres. ApJ, 556 (2001), pp. 872--884.\\
Alexander, D. R., \& Ferguson, J. W. Low-temperature Rosseland opacities. ApJ, 437 (1994), pp. 879--891.\\
Bodenheimer, P., Grossman, A. S., Decampli, W. M., Marcy, G. \& Pollack, J. B., Calculations of the evolution of the giant planets, Icarus 41 (1980), pp. 293--308.\\
Boss, A. P., Giant planet formation by gravitational instability, Science 276 (1997), pp. 1836--1839.\\
Boss, A. P. Testing Disk Instability Models for Giant Planet Formation, ApJ, 661 (2007), pp. L73--L76.\\
Boss, A. P., Wetherill, G. W.,  \& Haghighipour, N.,NOTE: Rapid Formation of Ice Giant Planets, Icarus, 156 (2002), pp. 291--295.\\
Decampli, W. M. and Cameron, A. G. W., Structure and evolution of isolated giant gaseous protoplanets, Icarus, 38 (1979), pp. 367-391.\\
Dominik, C., Blum, J., Cuzzi, J. N. \& Wurm, G., Growth of Dust as the Initial Step Toward Planet Formation. PPV, B. Reipurth, D. Jewitt, and K. Keil (eds.), University of Arizona Press, Tucson, (2007), 951, pp. 783--800.\\
Guillot, T., The composition of transiting giant extrasolar planets. Dans Physica Scripta - Physics of Planetary Systems, Nobel Symposium 135, Stockholm. (2007), eprint arXiv:0712.2500\\
Helled, R., Podolak, M. \& Kovetz, A., Planetesimal capture in the disk instability model. Icarus 185 (2006), pp. 64--71.\\
Helled, R., Podolak, M. \& Kovetz, A., Grain sedimentation in a giant gaseous protoplanet. Icarus (2008), in press. \\
Helling, Ch. \& Woitke, P., Dust in brown dwarfs. V. Growth and evaporation of dirty dust grains. Astron. Astrophys., 455 (2006), pp. 325--338.\\
Hubickyj, O., Bodenheimer, P., \& Lissauer, J. J. Accretion of the gaseous envelope of Jupiter around a 5Ð10 Earth-mass core, Icarus 179 (2005), pp. 415--431.\\
Lissauer, J. J. \& Stevenson, D. J., Formation of Giant Planets. PPV, B. Reipurth, D. Jewitt, and K. Keil (eds.), University of Arizona Press, Tucson (2007), 951, pp. 591--606.\\
Markiewicz, W. J., Mizuno, H., \& Voelk, H. J. Turbulence induced relative velocity between two grains, Astron. Astrophys. 242 (1991), pp. 286--289.\\
Mayer, L., Lufkin, G., Quinn, T.,  \& Wadsley, J. Fragmentation of Gravitationally Unstable Gaseous Protoplanetary Disks with Radiative Transfer, ApJ , 661 (2007),  pp. L77--L80.\\
Mayer, L., Quinn, T., Wadsley, J.,  \& Stadel, J. The Evolution of Gravitationally Unstable Protoplanetary Disks: Fragmentation and Possible Giant Planet Formation. ApJ , 609 (2004),  pp. 1045--1064.\\
Mayer, L., Quinn, T., Wadsley, J.,  \& Stadel, J. Formation of Giant Planets by Fragmentation of Protoplanetary Disks, Science, 298 (2002),  pp. 1756--1759.\\
Papaloizou, J. C. B., Nelson, R. P., Kley, W., Masset, F. S. \& Artymowicz, P. Disk-Planet Interactions During Planet Formation. PPV, B. Reipurth, D. Jewitt, and K. Keil (eds.), University of Arizona Press, Tucson, (2007), 951, pp. 655--668.\\
Podolak, M., Pollack, J. B., \& Reynolds, R. T. The interaction of planetesimals with protoplanetary atmospheres, Icarus 73 (1987), pp. 163--179.\\
Podolak, M. The contribution of small grains to the opacity of protoplanetary atmosphere, Icarus 165 (2003), pp. 428--437.\\
Pollack, J. B., McKay, C.P. \& Christofferson, B.M. A calculation of the Rosseland mean opacity of dust grains in primordial Solar System nebulae, Icarus 64 (1985), pp. 471--492.\\
Pollack, J. B., Hubickyj, O., Bodenheimer, P., Lissauer, J. J., Podolak, M., \& Greenzweig, Y. Formation of the giant planets by concurrent accretion of solids and gas, Icarus 124 (1996), pp. 62--85.\\
Rafikov, R. R. Convective Cooling and Fragmentation of Gravitationally Unstable Disks, ApJ, 662 (2007),  pp. 642--650.\\ 
Saumon, D., Chabrier, G. \&  van Horn, H.M.  An equation of state for low-mass stars and giant planets, Astrophys. J. Suppl. 99 (1995), pp. 713--741.\\
Saumon, D. \& Guillot, T., Shock compression of deuterium and the interiors of Jupiter and Saturn, Astrophys. J. 609 (2004), pp. 1170--1180.\\
Volk, H.J. , Jones, F. , Morfill G. \& Roser S., Collisions between grains in a turbulent gas, Astron. Astrophys. 85 (1980), pp. 316--325. \\
Weidenschilling, S., Evolution of grains in a turbulent solar nebula, Icarus 60 (1986), pp. 553--567.\\
Wetherill,G.W., Comparison of analytical and physical modeling of planetesimal accumulation, Icarus 88 (1990), pp. 336--354.\\
Woitke, P. \& Helling, Ch., Dust in brown dwarfs. II. The coupled problem of dust formation and sedimentation. Astron. Astrophys. 399 (2003), pp. 297--313.
}
\newpage

\begin{table}[h!]
\hspace{-1.5cm}
\centering
\begin{tabular}{||c||c||c|c|c|c|c||}
\hline
M/M$_J$ & $\rho_{surface}$ (g cm$^{-3}$) & $P_c$ (dynes cm$^{-2}$) & $T_c$ (K) & R (cm) & $T_{eff}$ (K)& L (erg s$^{-1}$)\\
\hline\hline
0.3 & 5.7$\times 10^{-11}$ & 7.3$\times 10^2$ & 2.2$\times 10^2$ &  3.5$\times 10^{12}$ & 18 & 9.2$\times 10^{26}$\\
\hline
1 & 2.3$\times 10^{-11}$ & 1.0$\times 10^2$  & 4.8$\times 10^2$  & 5.3$\times 10^{12}$ & 28 & 1.3$\times 10^{28}$\\
\hline
3  & 1.3$\times 10^{-11}$ & 1.56$\times 10^3$ & 9.0$\times 10^2$ & 7.8$\times 10^{12}$ & 38 & 9.3$\times 10^{28}$\\
\hline
5  & 1.0$\times 10^{-11}$ & 3.2$\times 10^3$ & 1.4$\times 10^3$ & 8.4$\times 10^{12}$ & 46 & 2.1$\times 10^{29}$\\
\hline
7  & 7.6$\times 10^{-12}$  & 5.3$\times 10^3$  &  1.8$\times 10^3$ & 9.1$\times 10^{12}$  & 52 & 4.4$\times 10^{29}$\\
\hline
10  & 5.1$\times 10^{-12}$ & 1.1$\times 10^4$  & 2.0$\times 10^3$ & 1.1$\times 10^{13}$ & 61 & 1.2$\times 10^{30}$\\
\hline
\end{tabular}
\caption{{\small Physical parameters of the initial states}} \label{tab:3}
\end{table}

 \begin{table}[h!]
\centering
\begin{tabular}{||c|c|c|c|c|c|c||}
\hline
M/M$_J$ & Available time & Grain Mass & Core Mass of  & Core Mass of & Core Mass of & $M_{core}/M_{grains}$\\[-0.13 in]
&(years)&(M$_{\oplus}$)&0.01 cm grains &0.1 cm grains&1 cm grains&\\
\hline\hline
0.3 &1.2$\times 10^6$& 0.63& 0.51 M$_{\oplus}$ & 0.53 M$_{\oplus}$ &  0.55M$_{\oplus}$ & $\sim$81-87\%\\
\hline
1 & 1.5$\times 10^5$ &2.1&0.26 M$_{\oplus}$  &0.25 M$_{\oplus}$ & 0.25M$_{\oplus}$ &$\sim$12\%\\
\hline
3  & 1.6$\times 10^4$ & 6.4& 0.81M$_{\oplus}$ &0.77M$_{\oplus}$ & 0.79M$_{\oplus}$ &$\sim$12-13\%\\
\hline
5  &  0 & 10.6 & No core  & No core   & No core   & 0\% \\
\hline
7  &  0 & 14.8&No core  &  No core  & No core & 0\%\\
\hline
10  & 0 &21.2&No core  & No core & No core & 0\%\\
\hline
\end{tabular}
\caption{{\small Core mass for different planetary masses. Column 2 lists the time required to reach a central temperature of $\sim$ 1300 K, a temperature high enough to evaporate the silicate grains. Column 3 gives the available grain mass. Columns 4-6 give the core mass for initial grain sizes (radii) of 0.01 to 1 cm, respectively. The last column gives the percentage of the core mass in comparison to the total grain mass in the protoplanet. }} \label{tab:3}
\end{table}

\newpage
\begin{figure}[h!]
    \centering
    \includegraphics[width=7in]{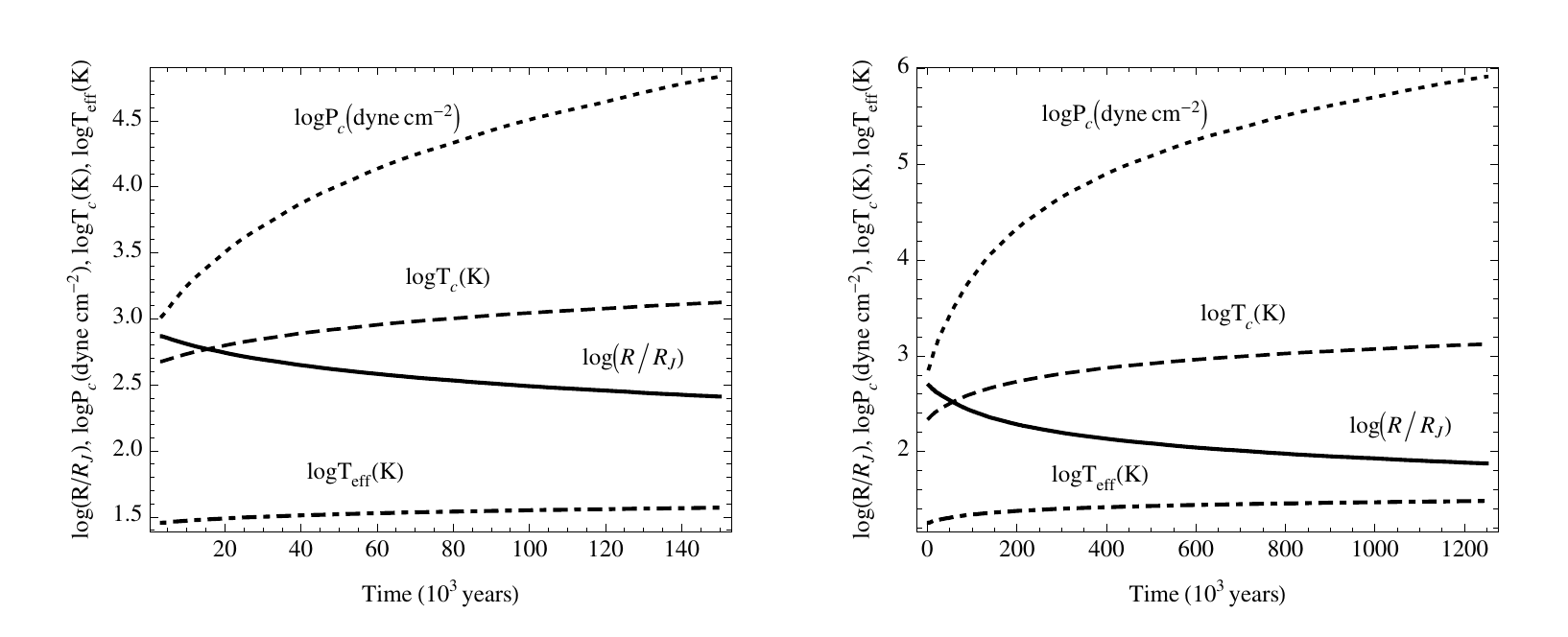}
    \caption[err]{Central pressure, central temperature, radius and effective temperature of Jupiter (left) and Saturn (right) mass protoplanets as a function of time.}
\end{figure}

\begin{figure}[h!]
    \centering
    \includegraphics[width=5.5in]{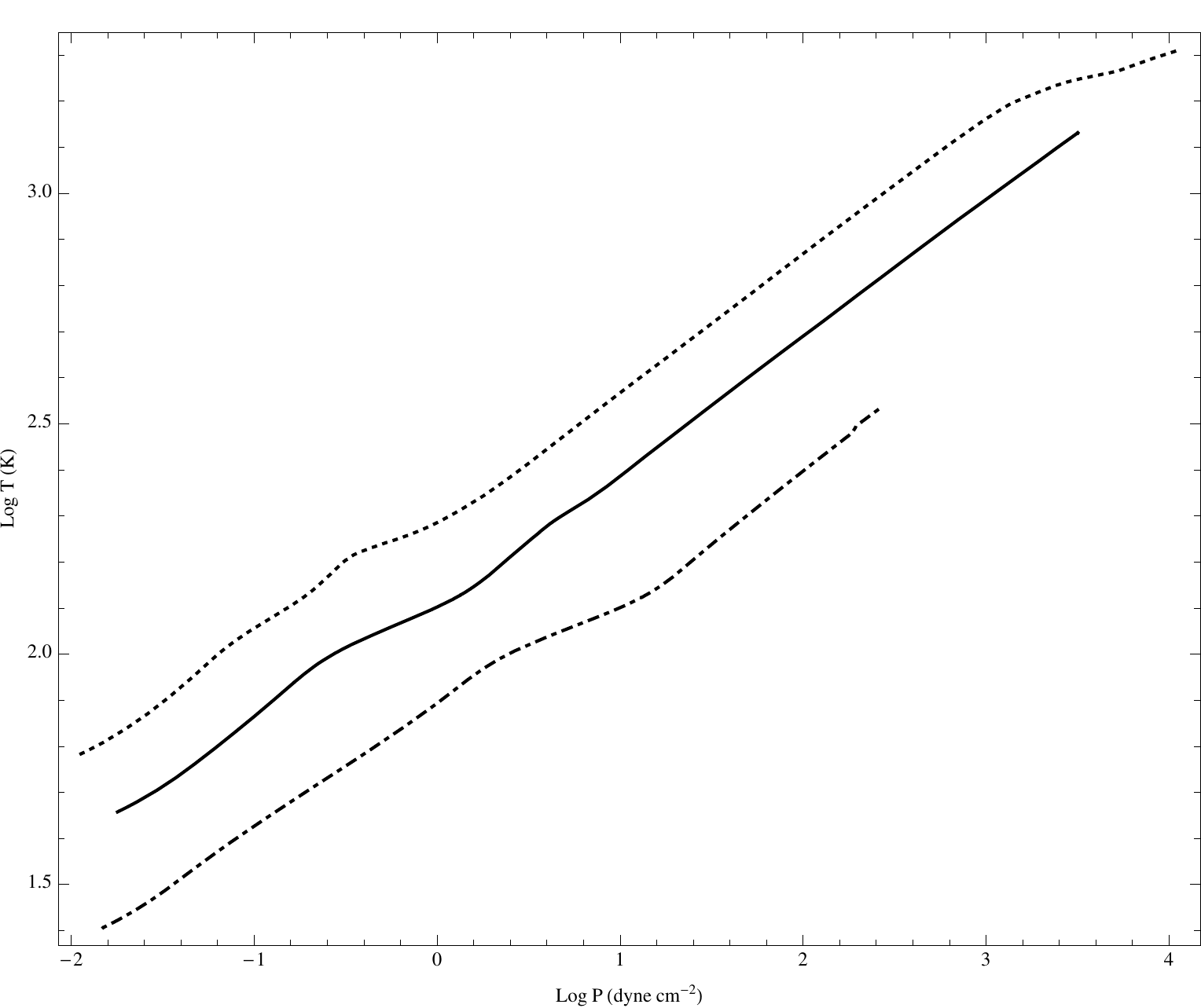}
    \caption[err]{Initial temperature-pressure profiles.  The dashed-dotted curve represents a Jupiter mass clump while the solid and dotted curves show the initial configuration for objects with 5 and 10 Jupiter masses, respectively.}
\end{figure}

\begin{figure}[h!]
    \centering
    \includegraphics[width=7in]{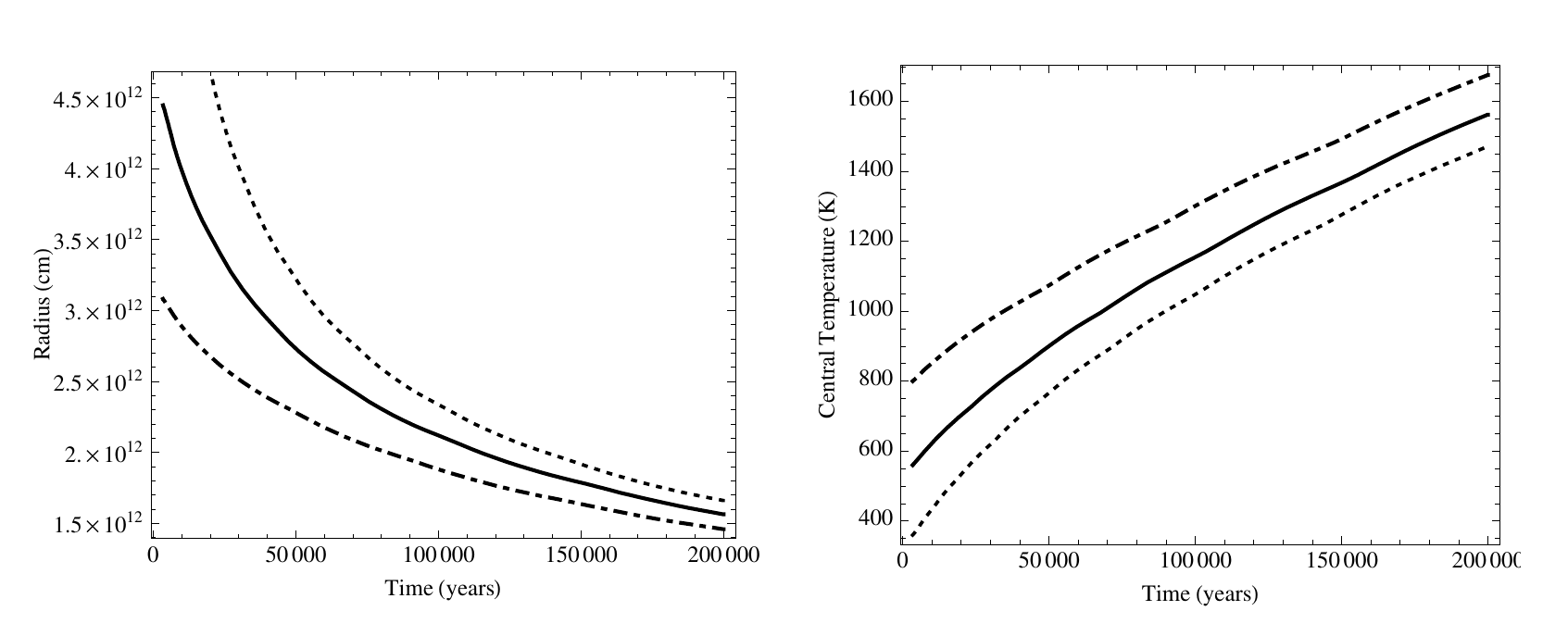}
    \caption[err]{Evolution of a Jupiter mass object with different initial conditions. Left: Radius as a function of time. Right: Central temperature as a function of time.
    The dotted curve represents a Jupiter mass object with an initial radius of $\sim$ 0.5 AU and a central temperature of 360 K. The solid and dashed-dotted curves show the evolution of Jupiter mass objects with initial radii of $\sim$ 0.3 and 0.2 AU and central temperatures of 560 and 800 K, respectively.}
\end{figure}

\end{document}